\documentclass[a4paper,12pt]{article}

\usepackage{indentfirst}
\usepackage{amsfonts,amsmath,amssymb,bm,a4wide,graphicx,cite,makeidx,multicol,color}

\begin{document}

\title{Neutrino spin operator and dispersion \\ in moving matter}

\author{A.Grigoriev\footnote{ax.grigoriev@gmail.com}, \ A.Studenikin, \ A.Ternov
   }

\date{}
\maketitle

\begin{abstract}
We found a spin integral of motion for neutrino, which propagates in moving and polarized matter. The operator obtained opens up the possibility of consistent classification of neutrino states in such a medium and, as a consequence, a systematic description of the related physical phenomena. Using the operator, we obtain a dispersion relation for neutrinos and consider its particular cases.
\end{abstract}

The problem of describing the neutrino motion in extreme external conditions (intense fields and dense matter) has achieved exceptional relevance in the multi-messenger era \cite{Spurio-Book:2015}, when humanity is able to register signals of various nature coming from the Universe.

The study of the structure of neutrino signal sources -- intense astrophysical processes involving compact objects (type II supernovae, gamma-ray bursts) -- within the framework of the planned neutrino mega-projects \cite{ShuLiZhou_PRD2016} requires a detailed understanding of the evolution of a neutrino beam inside these objects or in their environment. At that, evolution is not limited only to spatial motion, but can also include time dependence of the flavor composition and helicity (flavor and spin-flavor oscillations). In general, evolution can be very complex, since it can be influenced by many factors connected with the interaction of neutrinos with the environment (dense medium and/or external fields) \cite{Mirizzi_NuovCim2016}. To consistently take these factors into account, a detailed development of the theory of neutrino motion in external environment is needed.

The propagation of a neutrino beam under external conditions in the simplest model cases can be analytically described by using the corresponding wave equation, from which a neutrino dispersion relation follows. Within the framework of this approach, it is possible to rigorously substantiate and reproduce the known results on the flavor and spin-flavor neutrinos oscillations in the medium and the electromagnetic field \cite{NotzRaff_NPB307,GiunStud_2015}, and also to obtain new ones \cite{GrigSavochStud_2007,TokStud_2014,PustStud_PRD2018}.

For instance, solutions of the so-called modified Dirac equation for neutrinos in a homogeneous medium and the corresponding expression for energy, from which, in particular, the well-known effect of neutrino spin oscillations follows, become helicity-dependent \cite{StudTern_PLB2005}. This feature can lead to various physical effects in matter, for example, to the emission of a photon (the spin light of neutrino \cite{StudTern_PLB2005,LobStud_PLB564,Lob_PLB619}), as well as to self-polarization of a neutrino beam \cite{LobStud_PLB2004}.

A rigorous derivation of these effects (calculation of processes amplitudes) is possible only on the basis of the classification of the initial and final neutrino states according to the full set of observables, including spin quantum numbers. This example recalls that the most important point in such problems, as in the well-known case of charged particle motion in the electromagnetic field \cite{Sokolov-Ternov-Rel-El}, is the separation of states according to the spin quantum number, i.e. within the consistent approach, the exact solving of the Dirac equation must involve determination of the independent integral of motion - the spin operator.

In this paper, we study \emph{the modified Dirac equation} that describes a neutrino coherently interacting with particles of the external matter with taking into account possible effects of matter motion and polarization. In general case, this equation can be written as follows:
\begin{equation}\label{Dirac_eq}
  \left\{ i\gamma_{\mu}\partial^{\mu}- \frac{1}{2}\gamma_{\mu}(1+\gamma_5)f^{\mu} - m \right\}\Psi(x)=0,
\end{equation}
where $f^{\mu}$ is the constant 4-vector, represented by the linear combination of the matter current vector $j^{\mu}$ and the polarization vector $\lambda^{\mu}$ with coefficients, determined by the type of neutrino interaction with matter particles (see, for instance, \cite{StudTern_PLB2005,Lob_PLB619}), $\gamma^{5}=-i\gamma^{0}\gamma^{1}\gamma^{2}\gamma^{3}$.

For the medium consisting of identical particles, the  current 4\nobreakdash-vector is defined as $j^{\mu}=n_0v^{\mu}=(n,n\mathbf{v})$, where the matter density $n$ is related to its velocity $\mathbf{v}$ and density $n_0$ in its rest frame by the standard kinematical relation $n=\gamma n_0$, where $\gamma=1/\sqrt{1-\mathrm{v}^2}$ is the Lorentz gamma factor. The polarization 4\nobreakdash-vector in the rest frame of the medium has the components $\lambda^{\mu}=(0,n_0{\bm \zeta})$, where the three-dimensional vector $\bm \zeta$ is the average value of the matter polarization vector \cite{LobStud_PLB2001}.

Matter polarization occurs due to the interaction with external magnetic field (see, for example, \cite{Nunok-Semik-Smir-Val:97}) and in this connection it is often assumed that equation (\ref{Dirac_eq}) describes scattering of neutrinos on  medium particles polarized by an external field. Note that there is another physical interpretation of this phenomenon, according to which a direct interaction of the neutrino \emph{induced magnetic moment} (IMM) with an external magnetic field takes place \cite{GrigKupchTern_PRB2019}. Both approaches are equivalent and lead to the same results.

In what follows, we will consider the problem of a massive electron neutrino propagation in moving homogeneous unpolarized medium consisting of electrons \footnote{The results of our work remain valid in the general case of a polarized medium also.}. Then the 4-vector $f^{\mu}$ is determined only by the matter current, and within the framework of the Minimally Extended Standard Model we have \cite{StudTern_PLB2005}:
\begin{equation}\label{f_mu}
  \frac{1}{2}\, f^{\mu}=\tilde{n}_0v^{\mu}, \quad \tilde{n}_0=\frac{1}{2\sqrt{2}} G_F (1+4\sin^2\theta_W)n_0,
\end{equation}
where the dependent on the type of interaction coefficient is included in the definition of the ``density with a tilde'' $\tilde{n}_0$.
Note that, the equation of the same type as (\ref{Dirac_eq}) but in a more general form, also appears in phenomenological theories that go beyond the Standard Model (in particular, in theories with Lorentz invariance violation \cite{CollKostel_1997} and in theories with non-standard neutrino interactions \cite{Ohlsson_2013}).

In the phenomenological respect, the problem of neutrino propagation in moving matter was  for the first time considered in  \cite{Stud_PAN2004}, where, on the basis of a semiclassical approach, the effect of neutrino spin flip in the transverse matter current was described and the related phenomenon of spin oscillations was discussed. This effect and its possible contribution to the evolution of neutrinos in astrophysical media were discussed later in \cite{VlasFuller_PLB2015,Volpe_IJMP2015,KartRaffVog_PRD2015,DobrKartRaff_PRD2016,Tian_PRD2017}. The theory of the phenomenon was further developed in \cite{PustStud_PRD2018}, where it was considered on the quantum level by calculating the amplitudes that enter the effective neutrino evolution equation. Neutrino spin oscillations in a supernova environment associated with the presence of IMM were considered in \cite{GrigKupchTern_PRB2019}, see also \cite{Tern_JETP2016,Tern_PRD2016}.

As it was indicated above, in order to consistently describe neutrino processes under various external conditions and to classify solutions of the corresponding equation, it is necessary to establish the form of the spin operator, which would be the integral of motion, and also to determine the dispersion relation that would depend on the spin quantum number.

Of the works listed above, the solution of the modified Dirac equation of the considered type with the establishment of the required spin operator was carried out only in \cite{Tern_JETP2016,Tern_PRD2016}. However, in these works, a particular form of the equation was investigated, and the final dispersion relation was obtained in the linear-field approximation. Note also that the particular case considered there corresponds to the null zero component $f^{0}$ of the 4-vector $f^{\mu}$. In our setting (\ref{f_mu}), this component cannot vanish.

Note also that related questions about the neutrino propagator in a moving medium were studied in \cite{PivovStud_2006,KaloshVoron_2019}, where the exact dispersion relation for these conditions was obtained in different representations. It reduces to a trivially unsolvable fourth-order equation for energy. The dispersion relation was also obtained in \cite{CollKostel_1997} for a more general equation. In these cases, the polarization properties of the particle were either not studied at all, or were not determined in closed form through the eigenvalues of the conserved spin operator, and thus the physical basis for the classification of the solutions remained unclear. In this study, we aimed to obtain the exact spin integral of motion for equation (\ref{Dirac_eq}) in an explicit form and to determine the dispersion relation with the corresponding quantum number that has a clear physical meaning.

By this means, it is necessary to find a spin operator commuting with the Hamiltonian of equation (\ref{Dirac_eq}), which has the form
\begin{equation}\label{H}
  \mathrm{H}=({\bm \alpha}{\mathbf{p}})-{\tilde{n}}({\bm \alpha}{\mathbf{v}})+\tilde{n}({\bm \Sigma}{\mathbf{v}})+\gamma^5\tilde{n}+\tilde{n}+\gamma^0m.
\end{equation}
In expression (\ref{H}), we have introduced the three-dimensional neutrino momentum $\mathbf p$ and used the notations $\alpha_i=\gamma^0\gamma^i$, $\Sigma_i=\gamma^0\gamma^5\gamma^i$,  $\tilde{n}=\gamma\tilde{n}_0$. As a basis for the required operator, we use the 4-vector spin polarization operator $\mathrm{T}^{\mu}$ (see, for instance, \cite{Sokolov-Ternov-Rel-El,Bordovitsyn:2002})
\begin{equation}\label{T_mu}
  \mathrm{T}^{\mu}=\gamma^5\gamma^{\mu}-\gamma^5p^{\mu}/m.
\end{equation}
In this operator, let us apply the momentum ``extension'' $p^{\mu} \rightarrow \tilde{p}^{\mu}\equiv p^{\mu}-f^{\mu}=p^{\mu}-\tilde{n}_0 v^{\mu}$ (herewith $\mathrm{T}^{\mu} \rightarrow \widetilde{\mathrm{T}}^{\mu}$) and compose a scalar product of the vector $\widetilde{\mathrm{T}}^{\mu}$ with the 4-vector $v^{\mu}$. It is easy to verify that the commutation of the obtained quantity $\widetilde{\mathrm{T}}^{\mu}v_{\mu}\equiv (\widetilde{\mathrm{T}}v)$ with the Hamiltonian (\ref{H}) is achieved on solutions of the Dirac equation. Then we define the spin operator by multiplying this quantity by $m$:
\begin{equation}\label{Tv}
  \mathrm{S}=m(\widetilde{\mathrm{T}}v).
\end{equation}
Introducing the notation $\widetilde{\mathrm{H}}=\tilde{p}^0=\mathrm{H}-\tilde{n}$, let us also present the operator $\mathrm{S}$ in the expanded form:
\begin{equation}\label{S_exp}
  \mathrm{S}=\gamma \left[\gamma^5\gamma^0 m-\gamma^5(\widetilde{\mathrm{H}}-\left(\tilde{{\mathbf{p}}}{\mathbf{v}})\right) -m\gamma^0({\bm \Sigma}{\mathbf{v}}) \right].
\end{equation}

Since the expression for the conserved spin operator in the form of a Lorentz scalar, it is also suitable to combine the rest of the kinematical characteristics -- energy and momentum -- into a quantity of this kind. Therefore, we compose a scalar combination
\begin{equation}\label{P}
  P = (\tilde{p}^{\mu}v_{\mu}) = (\tilde{p}v)=({p}v)-\tilde{n}_0, \ \ \tilde{n}_0=\tilde{n}/\gamma,
\end{equation}
which turns out to be convenient for writing kinematic quantities in this problem. The block form of the spin operator has a compact form then (here we use the standard representation of the gamma matrices and take into account the implementation of the operator $\widetilde{\mathrm{H}}\rightarrow \widetilde{E}=E-\tilde{n}$ on solutions of the Dirac equation):
\begin{equation}\label{S_block}
  \mathrm{S}=\begin{pmatrix}
      -\gamma m ({\bm \sigma}{\bm v}) & \gamma m+P \\
      -\gamma m+P & \gamma m ({\bm \sigma}{\bm v}) \\
    \end{pmatrix},
\end{equation}
where $\sigma_i$ are Pauli matrices. The eigenvalues of the operator $\mathrm{S}$ are easily found and have the form 
\begin{equation}\label{Lambda}
   s\sqrt{P^2-m^2} \equiv s\Lambda, \ \ s=\pm 1.
\end{equation}

The operator obtained characterizes the longitudinal polarization of the particle with respect to the 4-vector of the medium velocity $v^{\mu}$ and determines the stationary spin states of the particle $\Psi_s$, corresponding to the conserved quantity $\pm\Lambda$:
\begin{equation}\label{Eigen_S}
  \mathrm{S}\Psi_s=s\Lambda\Psi_s.
\end{equation}
In this regard, we note that the operator used in \cite{KaloshVoron_2019} to describe the neutrino polarization properties in the same problem, in terms of this article appears to be proportional to the scalar product $({\mathrm{T}}v)$, where no ``extension'' for the momentum is implemented in the 4-vector $\mathrm{T}^{\mu}$. This turns out to be sufficient for this value not to be an integral of motion (as in fact was noted in \cite{KaloshVoron_2019}).

Relation (\ref{Eigen_S}) together with the original equation (\ref{Dirac_eq}) allows us to determine the dispersion relation for neutrinos in a moving medium taking into account the spin quantum number $s$. To find it, one can perform a standard procedure and represent the wave function through two-component spinors $\Psi=(\varphi, \chi)^{\tau}$. Rewriting the original Dirac equation (\ref{Dirac_eq}) in the form of a system of equations for these spinors, we obtain the compatibility condition for the equations in the form
\begin{equation}\label{Dispersion}
  p^2-m^2=2\tilde{n}_0(P-s\Lambda),
\end{equation}
where $p^2=E^2-{\mathbf{p}}^2$. This dispersion relation is reduced to a fourth-order algebraic equation for the particle energy.

First of all, it should be noted that the analytical expression for the dispersion law can be written in various forms. Indeed, in \cite{KaloshVoron_2019} it was obtained in the form (entries are written in the notation of this work)
\begin{equation}\label{Dispersion_Kaloshin}
  p^2-m^2=2\tilde{n}_0\Big((pv)-s'\sqrt{(pv)^2-p^2}\Big),
\end{equation}
and is not reduced to (\ref{Dispersion}) by simple algebraic transformation. The quantity $s'$ also equals to $\pm 1$, but it should have a different physical meaning. However, when reduced to forth-order algebraic equation with respect to $E$, equations (\ref{Dispersion}) and (\ref{Dispersion_Kaloshin}) become a general dispersion relation that does not contain the spin quantum number and that was obtained in \cite{PivovStud_2006}:
\begin{equation}\label{Dispersion_PivovStud}
  (p^2-m^2)^2-4\tilde{n}_0(pv)(p^2-m^2)+4\tilde{n}_0^2p^2=0.
\end{equation}
We also note one more representation of the dispersion relation (obtained directly from (\ref{Dispersion})) - completely through the ``extended'' values:
\begin{equation}\label{Dispersion_Tilde}
  \widetilde{p}^2-m^2=-\tilde{n}_0(2s\Lambda+\tilde{n}_0).
\end{equation}

Let us consider some special cases of the dispersion relation solution. For a medium at rest (when ${\mathbf v}=0)$, we have  $P=E-\tilde{n}_0=\widetilde{E}$, $\tilde{p}^2=\widetilde{E}^2-{\mathbf{p}}^2$. Next, we will use the relation (\ref{Dispersion_Tilde}), rewriting it in the form 
\begin{equation}
  (\sqrt{\widetilde{E}^2-m^2}+s\tilde{n}_0)^2=p^2,
\end{equation}
whence, introducing the ``energy sign'' $\varepsilon=\pm 1$, we obtain the previously known result \cite{StudTern_PLB2005} ($\mathrm{p}=|\mathbf{p}|$):
\begin{equation}\label{E_v=0}
  E=\varepsilon\sqrt{(\mathrm{p}-s\tilde{n}_0)^2+m^2}+\tilde{n}_0.
\end{equation}

In a similar way, a solution can be obtained for the case of the neutrino and matter parallel motion ($\mathbf{v} {\parallel} \mathbf{p}$). The expression for the energy in this case has the form
\begin{equation}\label{E_v||p}
  E=\varepsilon\sqrt{(\mathrm{p}\mp\mathrm{v}\tilde{n}-s\tilde{n})^2+m^2}\pm s\mathrm{v}\tilde{n}+\tilde{n},
\end{equation}
where the upper signs correspond to the motion of the medium along the direction of neutrino propagation, and the lower ones correspond to the opposite.

Note that in expressions (\ref{E_v=0}), (\ref{E_v||p}) the quantum number $s$ is the helicity of the particle (see \cite{StudTern_PLB2005}), and they can be used for determination of the energy difference between ultrarelativistic left- and right-handed neutrinos $\Delta E=E_L-E_R$ in the problem of neutrino oscillations in matter \cite{GrigStudTern_NuclPhys2006}. For the case of non-moving matter, from equation (\ref{E_v=0}) the standard expression follows known from the theory of neutrino oscillations in matter $\Delta E=2\tilde{n}_0=\frac{1}{\sqrt{2}} G_F (1+4\sin^2\theta_W)n_0$ \cite{LimMarc_PRD88}. Expression (\ref{E_v||p}) in the case of neutrino motion along the matter flow gives:
$$\Delta E=2\tilde{n}_0\sqrt{\frac{1-\mathrm{v}}{1+\mathrm{v}}},$$
and in the opposite direction: 
$$\Delta E=2\tilde{n}_0\sqrt{\frac{1+\mathrm{v}}{1-\mathrm{v}}}.$$
Both expressions agree with the general formula, which takes into account arbitrary directions of matter velocity and polarization, and which was previously found within the framework of the semiphenomenological approach \cite{GrigLobStud2002}.

In the case of transversal matter motion, $(\mathbf{p v})=0$, the dispersion relation leads to a fourth-order equation with respect to $E$. For an ultrarelativistic neutrino, with the momentum being the largest parameter, i.e. $\mathrm{p} \gg \tilde{n}$, $\mathrm{p} \gg m$, its solutions corresponding to $\varepsilon=1$ can be approximately represented as
\begin{align}
      E_{s=+1} &=\sqrt{\big(\mathrm{p}-\tilde{n}(1-\mathrm{v}^2)\big)^2+m^2}+\tilde{n}(1-\mathrm{v}^2),\notag \\
      E_{s=-1} &=\sqrt{\big(\mathrm{p}+\tilde{n}(1-\mathrm{v}^2)\big)^2+4\tilde{n}^2\mathrm{v}^2+m^2}+\tilde{n}(1+\mathrm{v}^2).\label{E_p_perp_v}
\end{align}
Under the condition $\mathrm{v}\rightarrow0$ they reduces to (\ref{E_v=0}).

In conclusion, we recall that the considered formalism, based on the spin integral of motion, makes it possible to consistently describe  neutrino quantum states in matter with uniform and conserved characteristics -- density, velocity, and polarization. It can be applied to problems of relativistic neutrino astrophysics, where the dynamics of the neutrino spin plays an important role. The latter aspect may be key for the interpretation of the expected data on the registration of astrophysical neutrinos by large-volume detectors such as JUNO and Hyper-Kamiokande. Thus, one of the specific areas of our formalism application may be neutrino spin oscillations, which physics is being actively studied for various types of matter within the common problem of establishing the neutrino role in supernova explosions \cite{Janka2017}.

Another important possibility of application of the stationary neutrino quantum states in matter is the consistent description of various particle interaction processes participating neutrinos in dense astrophysical media. Due to emerging difference between two neutrino spin states (as can be seen, for example, from formulas (\ref{E_v=0}),(\ref{E_v||p}) and (\ref{E_p_perp_v})) such processes begin to depend on neutrino spin orientation. Given this fact, well-known processes can acquire new interesting properties, and in some cases there can take place the processes forbidden in vacuum. The relevant example is the spin light of neutrino in matter -- the emission of a photon by a neutrino moving in matter upon its transition between different spin states \cite{StudTern_PLB2005,LobStud_PLB564,Lob_PLB619}. The properties of this phenomenon turn out to be very sensitive to medium characteristics, including its motion. In dense astrophysical objects, the radiation can be intense enough to discuss the possibility of its experimental registration \cite{Gr-Lok-St-Ternov-JHEP:17}.

To conclude, the presented approach based on our spin integral of motion, is a new step in the consistent theoretical description of neutrino motion in a medium. It has specific areas of application in modern astrophysics and expands the possibilities of describing the elementary particles interaction processes in astrophysical conditions.

\end{document}